\documentclass[aps,prl,twocolumn,superscriptaddress,altaffilletter,nobibnotes,nofootinbib,10pt,showpacs,showkeys,preprintnumbers]{revtex4-1}

\usepackage{graphicx}
\usepackage{dcolumn}
\usepackage{bm}
\usepackage{lineno}
\usepackage{float}
\usepackage{color}
\usepackage{preprintcover}
\PreprintIdNumber{CERN-PH-EP-2012-212}
\PreprintCoverPaperTitle{Net--charge fluctuations\\ in Pb--Pb collisions at $\sqrt{s_{\rm NN}}= 2.76$~TeV}
\PreprintCoverAbstract{
We report the first measurement of the net--charge fluctuations in Pb--Pb collisions at 
$\sqrt{s_{\rm NN}}= 2.76$~TeV, measured with the ALICE detector at the CERN Large Hadron 
Collider. The dynamical fluctuations per unit entropy
are observed to decrease 
when going from peripheral to central collisions. An additional reduction in the
amount of fluctuations is seen in comparison to the results from lower energies. 
We examine the dependence of fluctuations on 
the pseudo--rapidity interval, which may account for the dilution of fluctuations
during the evolution of the system. 
We find that the fluctuations at LHC are smaller compared to the measurements at
the Relativistic heavy Ion Collider~(RHIC), and as such, closer to what has been
theoretically predicted for the formation of the Quark--Gluon Plasma~(QGP). 
}
\PreprintJournalName{PRL}
\newcommand{\pT}{$p_{\rm{T}}$~}
\newcommand{\dca}{$dca$}

\newcommand{\ITS}{\rm{ITS}}
\newcommand{\SPD}{\rm{SPD}}

\newcommand{\TPC}{\rm{TPC}}

\newcommand{\VZERO}{\rm{VZERO}}

\newcommand{\nudyn}{$\nu_{(+-,{\rm dyn})}$}
\newcommand{\nudyncorr}{$\nu^{\rm corr}_{(+-,{\rm dyn})}$}
\newcommand{\pt}{$p_{\rm{T}}$}
\newcommand{\deltaeta}{$\Delta \eta$}
\newcommand{\nch}{$\langle N_{\rm ch} \rangle$}
\newcommand{\ntotal}{$\langle N_{\rm total} \rangle$}
\newcommand{\nchnudyn}{$\langle N_{\rm ch} \rangle    \nu^{\rm corr}_{(+-,{\rm dyn})}$}

\newcounter{vers}

\definecolor{dgreen}{cmyk}{1.,0.,1.,0.2}        
\definecolor{orange}{cmyk}{0.,0.353,1.,0.}    

\begin{document}
\setcounter{page}{2}
\title{Net--Charge Fluctuations in Pb--Pb collisions at $\sqrt{s_{\rm NN}}= 2.76$~TeV}

\collaboration{ALICE Collaboration}

%
\author{B.~Abelev}
\affiliation{Lawrence Livermore National Laboratory, Livermore, California, United States}
\author{J.~Adam}
\affiliation{Faculty of Nuclear Sciences and Physical Engineering, Czech Technical University in Prague, Prague, Czech Republic}
\author{D.~Adamov\'{a}}
\affiliation{Nuclear Physics Institute, Academy of Sciences of the Czech Republic, \v{R}e\v{z} u Prahy, Czech Republic}
\author{A.M.~Adare}
\affiliation{Yale University, New Haven, Connecticut, United States}
\author{M.M.~Aggarwal}
\affiliation{Physics Department, Panjab University, Chandigarh, India}
\author{G.~Aglieri~Rinella}
\affiliation{European Organization for Nuclear Research (CERN), Geneva, Switzerland}
\author{A.G.~Agocs}
\affiliation{KFKI Research Institute for Particle and Nuclear Physics, Hungarian Academy of Sciences, Budapest, Hungary}
\author{A.~Agostinelli}
\affiliation{Dipartimento di Fisica dell'Universit\`{a} and Sezione INFN, Bologna, Italy}
\author{S.~Aguilar~Salazar}
\affiliation{Instituto de F\'{\i}sica, Universidad Nacional Aut\'{o}noma de M\'{e}xico, Mexico City, Mexico}
\author{Z.~Ahammed}
\affiliation{Variable Energy Cyclotron Centre, Kolkata, India}
% ...

\vspace{0.1cm}

\begin{abstract} 
We report the first measurement of the net--charge fluctuations in Pb--Pb collisions at 
$\sqrt{s_{\rm NN}}= 2.76$~TeV, measured with the ALICE detector at the CERN Large Hadron 
Collider. The dynamical fluctuations per unit entropy
are observed to decrease 
when going from peripheral to central collisions. An additional reduction in the
amount of fluctuations is seen in comparison to the results from lower energies. 
We examine the dependence of fluctuations on 
the pseudo--rapidity interval, which may account for the dilution of fluctuations
during the evolution of the system. 
We find that the fluctuations at LHC are smaller compared to the measurements at
the Relativistic heavy Ion Collider~(RHIC), and as such, closer to
what has been theoretically predicted for the formation of the Quark--Gluon Plasma~(QGP). 
\end{abstract}
\pacs{25.75.-q,25.75.Nq,12.38.Mh}
\maketitle

\cleardoublepage
\newpage
The ALICE 
experiment~\cite{Aamodt08} at the Large Hadron Collider (LHC)
is a multi--purpose detector designed to study the formation and 
evolution of nuclear matter at high temperatures and energy 
densities. One of the major goals of the experiment is to explore 
as many signals as possible towards characterizing the properties of
the Quark--Gluon Plasma (QGP),
the deconfined 
state of quarks and gluons, produced in high energy heavy--ion collisions. 
The study of event--by--event fluctuations provides a powerful tool
to characterize the thermodynamic properties of the system.
The fluctuations of conserved quantities in a finite phase space window, 
like net--charge of the system, are predicted to be one of the most sensitive signals of the QGP formation and
phase transition, and may provide complementary understanding of strong
interactions~\cite{JeonKoch00,asakawa,
dtilde,JeonKochReview,JeonKoch99,Asakawa00,Shuryak01,AbdelAziz05}.

In the QGP phase, the charge carriers are quarks with fractional charges,
whereas the particles in a hadron gas (HG) carry unit charge. 
The fluctuations in the net--charge depend on the squares of the charge states present in the
system. Consequently, the net--charge fluctuations in the QGP phase
are significantly smaller compared to that of a HG~\cite{JeonKoch00}.
At the same time, if the initial QGP phase is strongly gluon
 dominated, the fluctuation per entropy may further be reduced as the
hadronization of gluons increases the entropy~\cite{asakawa}.
Thus the net--charge
fluctuations are strongly dependent on which phase they originate from.
However, the net--charge fluctuations may get affected by uncertainties arising from 
volume fluctuations, so one considers the fluctuations of the
ratio, $R=N_+/N_-$.  Here  $N_+$ and $N_-$ 
are the numbers of positive and negative particles
respectively, measured in a specific transverse momentum (\pT) and 
pseudo--rapidity ($\eta$) window. The parameter $R$ is 
related to the fluctuations of the net--charge via the
$D$--measure as per the
following expression~\cite{JeonKoch00,dtilde,JeonKochReview}:
\begin{eqnarray}
D =  \langle N_{\rm ch} \rangle \langle \delta R^2 \rangle \approx 
4 \frac{\langle \delta Q^2\rangle}{\langle N_{\rm ch} \rangle},
\end{eqnarray}
\noindent 
which provides a measure of the charge fluctuations per unit entropy.
Here the $\langle ... \rangle $ denotes an average of 
the quantity over an ensemble of events.
The term $\langle \delta Q^2\rangle$ is the variance of net charge, $Q = N_+ -
N_-$ and $N_{\rm ch} = N_+ + N_-$. 
The $D$--measure has been estimated for several different theoretical
considerations including those of the lattice calculations. In a simple picture
by neglecting quark--quark interactions, $D$ is found to be
approximately 4 times smaller for a
QGP compared to a HG~\cite{JeonKoch00}.
Lattice calculations which include the quark--quark interactions give a quantitatively 
different estimate for a QGP phase, still significantly smaller than for a HG. 
It has been shown that $D=4$
for an uncorrelated pion gas, and after taking resonance yields into
account, the value decreases to $D\simeq 3$.
For a QGP, $D$ is significantly lower and has been calculated
to be $D\simeq1.0$--1.5~where the uncertainty arises from the uncertainty of relating 
the entropy to the number of charged particles in the final state~\cite{JeonKochReview}.
Thus, a measurement of $D$ can be effectively used as a probe for
distinguishing the two phases, the HG and the QGP.
However in reality, 
these fluctuations may get diluted in the rapidly expanding medium due
to diffusion of particles in rapidity space~\cite{Shuryak01,AbdelAziz05}. 
Several other effects, such as collision dynamics, radial flow,
resonance decays and final state
interactions may also affect the amount of measured fluctuations~\cite{JeonKoch00,Voloshin,zaranek,Abelev09}.

In the experiment, the net--charge
fluctuations are best studied~\cite{CERES,NA49,AdcoxPRL89,Adams03c,Abelev09,Pruneau02} 
by calculating the quantity \nudyn~defined as:
\begin{eqnarray}
\nu_{(+-,{\rm dyn})} = 
 \frac{\langle N_+(N_+-1) \rangle}{\langle N_+ \rangle ^2} +
 \frac{\langle N_-(N_--1) \rangle}{\langle N_- \rangle ^2} \nonumber \\
- 2\frac{\langle N_-N_+ \rangle}{\langle N_- \rangle \langle N_+ \rangle},
\end{eqnarray}
\noindent which is a measure of the relative correlation strength of
particle pairs. 
A negative value of \nudyn~signifies the dominant contribution from
correlations between pairs of opposite charges. On the other hand, a
positive value indicates the significance of the same charge pair
correlations. 
The \nudyn~has been found to be 
robust against random efficiency losses~\cite{Pruneau02,
  christiansen,Nystrand}.
$D$--measure and \nudyn~are related to each other by~\cite{JeonKochReview}:
\begin{eqnarray}
\langle N_{\rm ch} \rangle    \nu_{(+-,{\rm dyn})}  
\approx  D - 4.
\label{Eq:D}
\end{eqnarray}
The values of \nudyn~need to be corrected for global charge
conservation~\cite{Pruneau02}. 
The predictions for the $D$--measure are based on the assumption of 
vanishing net--charge in the system. However, in a realistic
situation, the system under consideration has a small but finite
net--charge. A correction due to finite net--charge effect also needs to be applied~\cite{dtilde}.


In this letter, we report the first measurements of net--charge
fluctuations, by calculating \nudyn~and the $D$--measure,
as a function of collision centrality in \mbox{Pb--Pb} collisions at 
$\sqrt{s_{\rm NN}}=2.76$ TeV at the LHC with the ALICE detector. We also make a comparison of the
experimental results to the theoretical predictions. 

Details of the ALICE experiment and its detectors can be found in 
\cite{Aamodt08}. For this analysis, we have used the Time Projection Chamber 
(\TPC)~\cite{TPC} to reconstruct charged particle tracks.  The detector 
provides a uniform acceptance with an almost constant tracking efficiency
of about 80\% in the analyzed phase space ($|\eta| < 0.8$ and 0.2~GeV/$c$
$<$ \pT $<$ 5~GeV/$c$). 
The interaction vertex was measured using
the Silicon Pixel Detector (\SPD), the innermost detector of 
the Inner Tracking System (\ITS). In 
the analysis, we have considered events with a vertex 
$|v_{\rm z}| < 10$~cm to ensure a uniform acceptance in the central pseudo--rapidity 
region. 
The minimum bias trigger consisted of a coincidence of at least
one hit on each of the two \VZERO~scintillator detectors, 
positioned on both sides of the interaction point, while at the
startup of data taking period an additional requirement of having a 
coincidence with a signal from the SPD was also introduced.
The background events coming from parasitic beam interactions are
removed by a standard offline event selection procedure, which requires
the  \VZERO~timing information and hits in the SPD.

We present the results as a function of centrality that reflects the collision geometry. 
The collision centrality is determined by cuts on the
\VZERO~multiplicity~\cite{ALICEcharged}. 
 A study 
based on Glauber model fits \cite{Glauber,miller,toia} to the multiplicity distribution 
in the region corresponding to 90\% of the most central collisions, where the 
vertex reconstruction is fully efficient, facilitates the determination of 
the cross section percentile and the number of participants. The
resolution in centrality is found to be $< 0.5$\% RMS for the most
central (0-5\%)
collisions, increasing towards 2\% RMS for peripheral (70-80\%)
collisions~\cite{ALICEcharged}.  

We require tracks in the \TPC~to have at least 80 reconstructed space 
points with a $\chi^2$ per \TPC~cluster of the momentum fit less than~4. 
We reject tracks with distance of closest approach (\dca) to 
the vertex larger than 3 cm both in the transverse plane and in the 
longitudinal direction. We have performed an alternative analysis with tracks 
reconstructed using the combined tracking of \ITS~and \TPC. In this case, the 
\dca~cuts were 0.3~cm in the transverse plane as well as in the longitudinal 
direction. The results obtained with both tracking approaches are in agreement. 

\begin{figure}
\begin{center}
\includegraphics[width=0.52\textwidth]{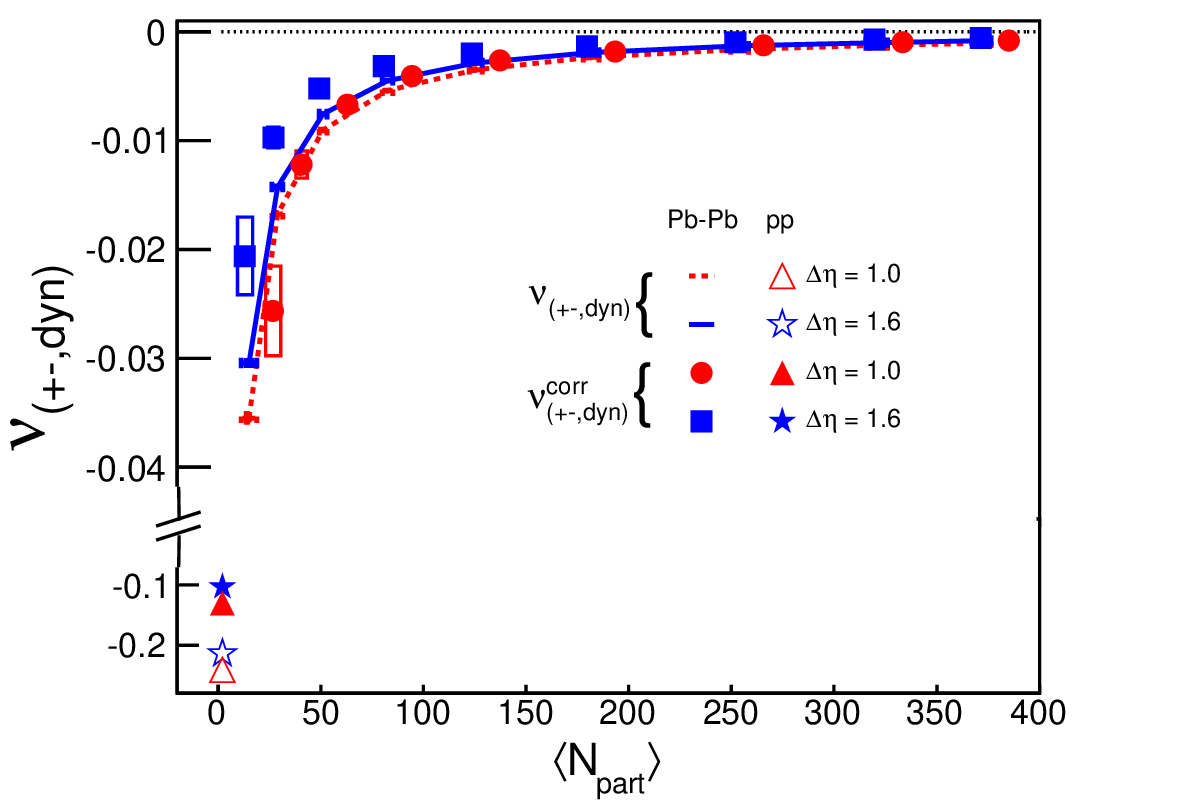}
\caption{
Dynamical net--charge fluctuations, \nudyn~and their corrected values, 
\nudyncorr, for charged particles produced in \mbox{Pb--Pb} collisions
at $\sqrt{s_{\rm NN}}=2.76$~TeV
as a function of centrality, expressed as the number of participating
nucleons. { {\nudyncorr~points are shifted along $x$-axis for better representation. }}
Superimposed are the results for \mbox{pp} collisions at $\sqrt{s}=2.76$~TeV.
The statistical (bar) and systematic (box) errors are plotted.
}
\label{fig:nudyn}
\end{center}
\end{figure}

The data analysis has been performed for  \mbox{Pb--Pb} collisions at
$\sqrt{s_{\rm NN}}=2.76$ TeV and \mbox{pp} collisions at
the same centre--of--mass energy.
An identical analysis procedure has been followed for 
both the data sets. 
We calculate the \nudyn~from the experimental measurements of positive 
and negative charged particles counted in \deltaeta~windows, defined 
around mid--rapidity (for example,  $\Delta \eta = 1$ corresponds 
to $-0.5 \le \eta \le 0.5$) and in the \pt~range from 0.2 to 5.0~GeV/$c$. 
{ {Consistency checks had been performed for
another \pT window, {\it viz.,} 0.3~GeV/$c$ $<$ \pT $<$ 1.5~GeV/$c$. The
differences in the fluctuation results are small, and included in the systematic errors.}}
In Figure~\ref{fig:nudyn}, we present the \nudyn~as a function of 
centrality, expressed in terms of the number of participating
nucleons. Moving from left to right along 
the $x$--axis of the figure corresponds to moving from peripheral to central collisions. 
The results are presented for $\Delta \eta = 1$
and 1.6, for both \mbox{Pb--Pb} and \mbox{pp} collisions.
In all cases, the magnitude of \nudyn~is observed to be negative,
indicating the dominance of the correlation term in Eq. 2. 
The absolute values of \nudyn~for \mbox{pp} collisions are larger
compared to those measured for \mbox{Pb--Pb} collisions.
When going from peripheral to central events, the absolute values of
\nudyn~are seen to decrease monotonically.

The values of \nudyn~have to be corrected for
global charge conservation and finite acceptance~\cite{Pruneau02}.
If all charges were accepted, the global 
charge conservation would lead to vanishing fluctuations. This will
yield the minimum value of
\nudyn~to be -4/$\langle N_{\rm total} \rangle$, where \ntotal~is
the average total number of
charged particles produced over full phase space.
The corrected \nudyn~is
\begin{eqnarray}
\nu^{\rm corr}_{(+-,{\rm dyn})} = \nu_{(+-,{\rm dyn})} +
\frac{4}{\langle  N_{\rm total} \rangle}.
\end{eqnarray}
The values of \ntotal~for \mbox{Pb--Pb} collisions have been
estimated from the experimental data~\cite{ALICE_ntotal}, whereas 
 for \mbox{pp} collisions, it is taken from
 PYTHIA~\cite{Ref:Pythia} event generator.  
As a reference, \nch~for \deltaeta=1 and \ntotal~values are 1637$\pm$61 and 
17165$\pm$772  for most central (0-5\%) \mbox{Pb--Pb} collisions, and 4.8$\pm$0.2 and 36.0
for \mbox{pp} collisions. These are systematic errors, statistical errors are negligible. 
The corrected values, \nudyncorr, are plotted in
Figure~\ref{fig:nudyn} as a function of the number of participating nucleons for \mbox{Pb--Pb}
and \mbox{pp} collisions. The absolute values of \nudyncorr are
smaller compared to \nudyn in all cases. The differences are more
apparent for \mbox{pp} and peripheral \mbox{Pb--Pb} collisions than
for central collisions. 

{{Taking the above corrections into account, we obtain, 
\begin{eqnarray}
D' = \langle N_{\rm ch} \rangle \nu^{\rm corr}_{(+-,{\rm dyn})}+4.
\end{eqnarray}
Alternatively, corrections to $D$-measure may also be
obtained using~\cite{dtilde}:
\begin{eqnarray}
D''=(\langle N_{\rm ch} \rangle \nu_{(+-,{\rm dyn})}+4)/(C_{\rm  \mu}C_{\rm \eta}) , 
\end{eqnarray}
where
$C_{\rm \mu}$ (=$\frac{{\langle N_{+}\rangle}^2}{{\langle N_{-} \rangle}^2}) $ corrects for the effects of the finite net charge,
and $C_{\rm \eta}$ (=1$-$$\frac{{\langle N_{\rm ch} \rangle}}{{\langle N_{\rm total}\rangle}})$ accounts for finite bin size in
rapidity as well as global charge conservation.
Differences in the two corrected values, $D'$ and $D''$, are within
4--9\%, depending on \deltaeta. Subsequently, mean values of $D'$ and $D''$ are
plotted in the figures as $D$, and differences of those have been included as
systematic errors.
}}

The systematic uncertainties have additional contributions from the following sources: 
(a) uncertainty in the determination of the interaction vertex, (b) different magnetic field polarities, 
(c) contamination from secondary tracks (\dca~cuts), (d) centrality definition using different detectors, 
(e) selection criteria at the track level,  (f) different tracking
scenarios, {{and (g) two different \pT windows.}}  
The total systematic error on \nudyncorr~amounts to 
6--10$\%$ in going from peripheral to central collisions.
The error on the product of number of charged particles and
\nudyncorr~remains within 7--13$\%$ at all centralities. 
The systematic and
statistical uncertainties in all the figures are 
represented by boxes and error bars, respectively. 
The statistical errors are small and 
within the sizes of the symbols in most cases. 

\begin{figure}
\begin{center}
\includegraphics[width=0.5\textwidth]{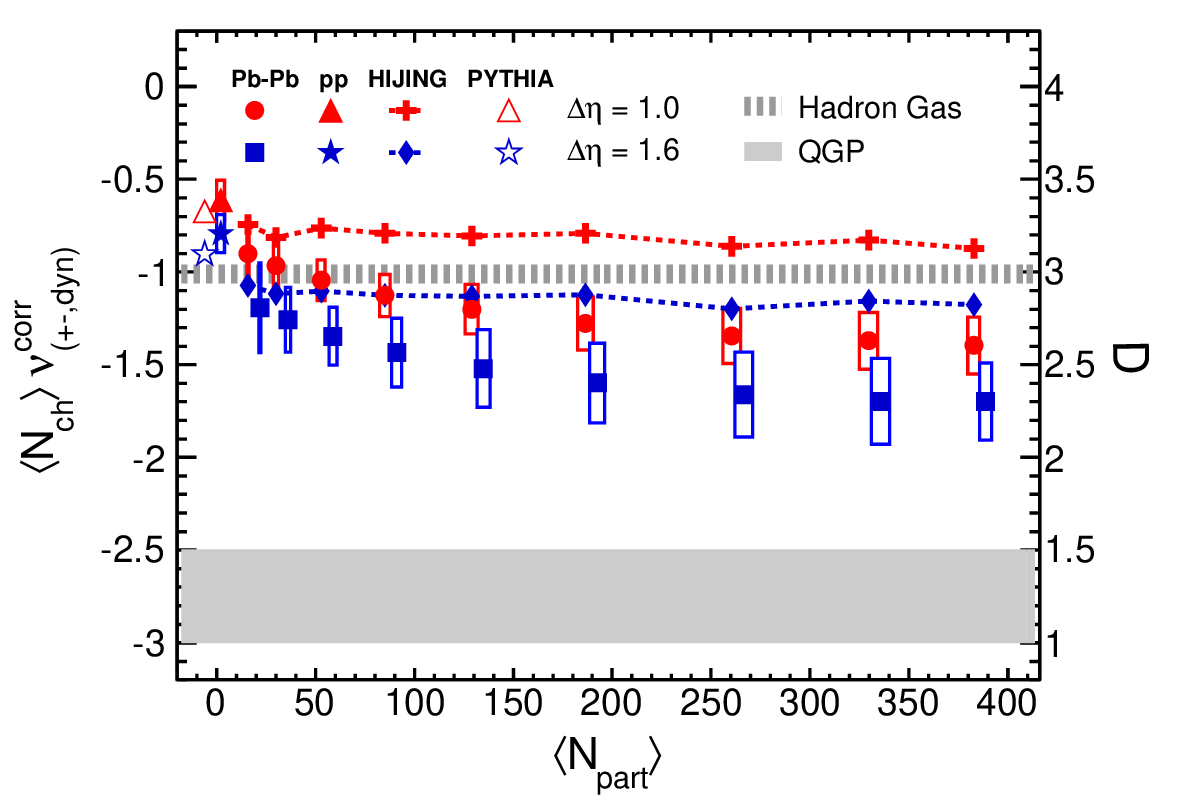}
\caption{ 
\nchnudyn~(left axis) and $D$ (right axis) as a function of the number of participants for
$\Delta \eta$=1 and 1.6 in \mbox{Pb--Pb} at
$\sqrt{s_{\rm NN}}=2.76$ TeV and pp collisions at $\sqrt{s}=2.76$ TeV.
Corresponding results from the HIJING {{and PYTHIA event
    generators are also presented.}}
The data points are shifted minimally along $x$-axis for clear view.
Both statistical (error bar) and systematic (box) errors are plotted.
}
\label{fig:fig_npart}
\end{center}
\end{figure}

\begin{figure}
\begin{center}
\includegraphics[width=0.49\textwidth]{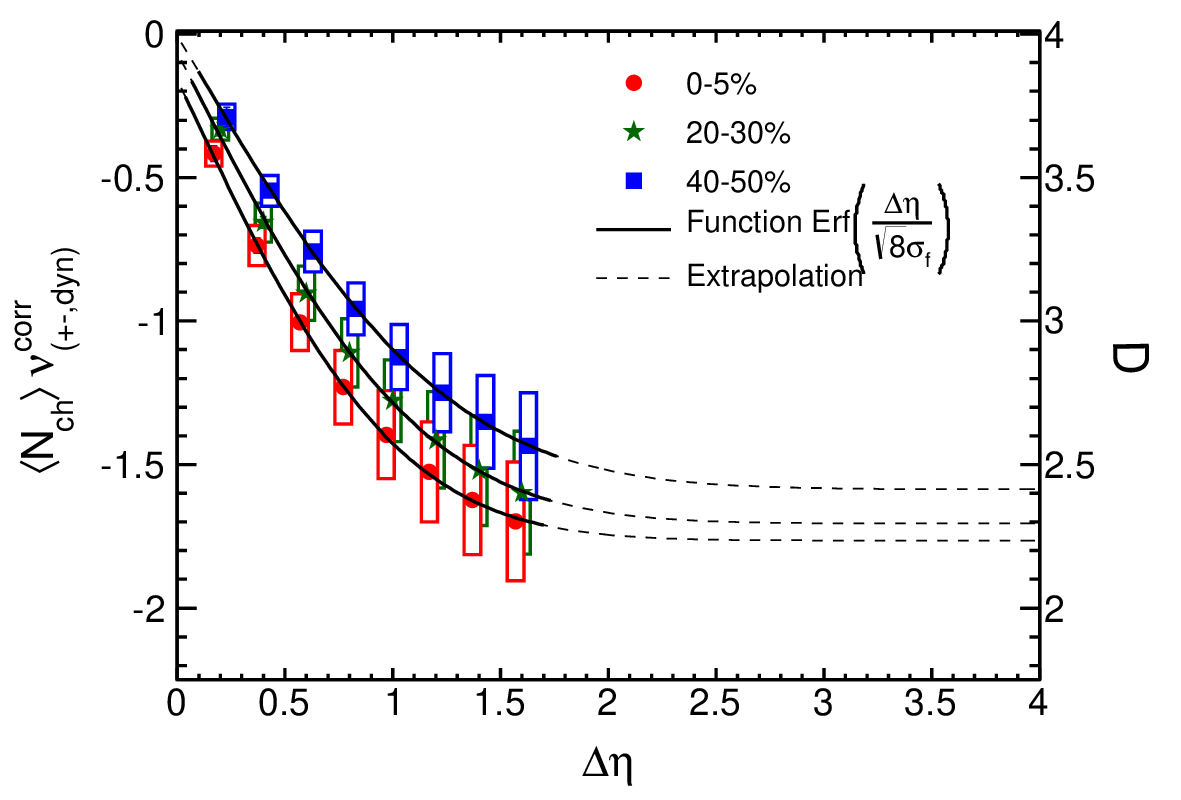}
\caption{
\nchnudyn~(left axis) and $D$~(right axis) as a function of 
\deltaeta~window for three different centrality bins in 
\mbox{Pb--Pb} collisions at $\sqrt{s_{\rm NN}}=2.76$ TeV. 
The data points are fitted with the functional form, 
erf($\Delta\eta/\sqrt{8}\sigma_{\rm f}$). The dashed lines correspond to the 
extrapolation of the fitted curves.  
The points are shifted minimally along $x$-axis for clear view.
Both statistical (error bar) and systematic (box) errors are shown.
}
\label{fig:D_eta}
\end{center}
\end{figure}

Figure~\ref{fig:fig_npart} 
presents the values of \nchnudyn and $D$ in the left and right axes,
respectively, as a function of the number of participating nucleons. 
The $\langle N_{\rm ch} \rangle$ values have been measured for different
centralities and \deltaeta~windows, and corrected for detector 
inefficiencies~\cite{ALICEcharged}. 
Both the results from the \mbox{Pb--Pb} and \mbox{pp} analyses are shown.
The shaded bands in the figure indicate the predictions for a HG
and a QGP. Results from the HIJING event generator~\cite{Ref:Hijing} at 
$\Delta \eta = 1$ and 1.6 are observed to be close to the HG line and at the same time
independent of centrality. 
The \mbox{pp} results agree well with the HG prediction. 
The experimental results for \mbox{Pb--Pb} for both the \deltaeta~windows 
are observed to be below the HG predictions and above those of the QGP. 
The values of $D$ for $\Delta \eta = 1.6$ are
lower compared to those for $\Delta \eta = 1$ for all centralities.

A decreasing trend of $D$ has been observed 
while going from peripheral to central collisions, as seen in Figure~\ref{fig:fig_npart}.
This centrality dependence may arise partly because of the presence of radial 
flow~\cite{Voloshin}. The radial 
flow velocity could lead to the kinetic focussing of 
the produced particles, causing a narrowing of the opening angles. 
This may affect the magnitude of net--charge fluctuations. 
The effect of radial flow on \nudyn~has been estimated by using an
afterburner~\cite{afterburner} on the HIJING events where the
particles get a boost in the transverse momenta because of the radial
flow velocity. 
We observe no significant difference between the
results from pure HIJING and HIJING with the afterburner. 
This indicates that the presence of radial flow may not be 
responsible for the centrality dependence of the $D$--measure.

The measured fluctuations may get diluted during the 
evolution of the system from hadronization 
to kinetic freeze-out because of the diffusion of charged hadrons in
rapidity. {{This has been addressed 
in refs.~\cite{Shuryak01,AbdelAziz05}, where 
a diffusion equation has been proposed 
to study the dependence of net--charge fluctuations
on the width of the rapidity window. 
Taking the dissipation into account, 
the asymptotic value of fluctuations may be close to the primordial
fluctuations.}}
This has been explored for the ALICE data points by 
plotting \nchnudyn~and $D$ as a function of \deltaeta~for three
centrality bins, as shown in Figure~\ref{fig:D_eta}.
We observe that for a given centrality bin, the
$D$--measure shows a strong decreasing trend with the
increase of \deltaeta. In fact, 
the curvature of $D$ has a decreasing slope with a flattening
tendency at large \deltaeta~windows.
Following the prescriptions of~\cite{Shuryak01,AbdelAziz05}, 
we fit the data points with the functional form,
erf($\Delta\eta/\sqrt{8}\sigma_{\rm f}$), 
which represents the diffusion in rapidity space.
Here, $\sigma_{\rm f}$ characterizes the diffusion at freeze--out. 
The resulting values of $\sigma_{\rm f}$ are $0.41 \pm 0.05$, 
$0.44 \pm 0.05$ and $0.48 \pm 0.07$ for the 0-5\%, 20-30\% and 40-50\%
centralities, respectively.
The fitted curves are shown as solid lines in Figure~\ref{fig:D_eta}.
The dashed lines are extrapolations of the
fitted curves to higher \deltaeta, which yield the asymptotic values of $D$.
For the top 5\% centrality, the measured values of $D$ are
$2.6 \pm 0.02 {\rm (stat.)} \pm 0.15 {\rm  (sys.)}$ for $\Delta \eta = 1$ and
$2.3 \pm 0.02 {\rm (stat.)} \pm 0.21 {\rm  (sys.)}$ for $\Delta \eta = 1.6$. 
The extrapolated value of $D$ is $2.24\pm 0.09{\rm (stat.)}\pm 0.21{\rm (sys.)}$. 

\begin{figure}
\begin{center}
\includegraphics[width=0.5\textwidth]{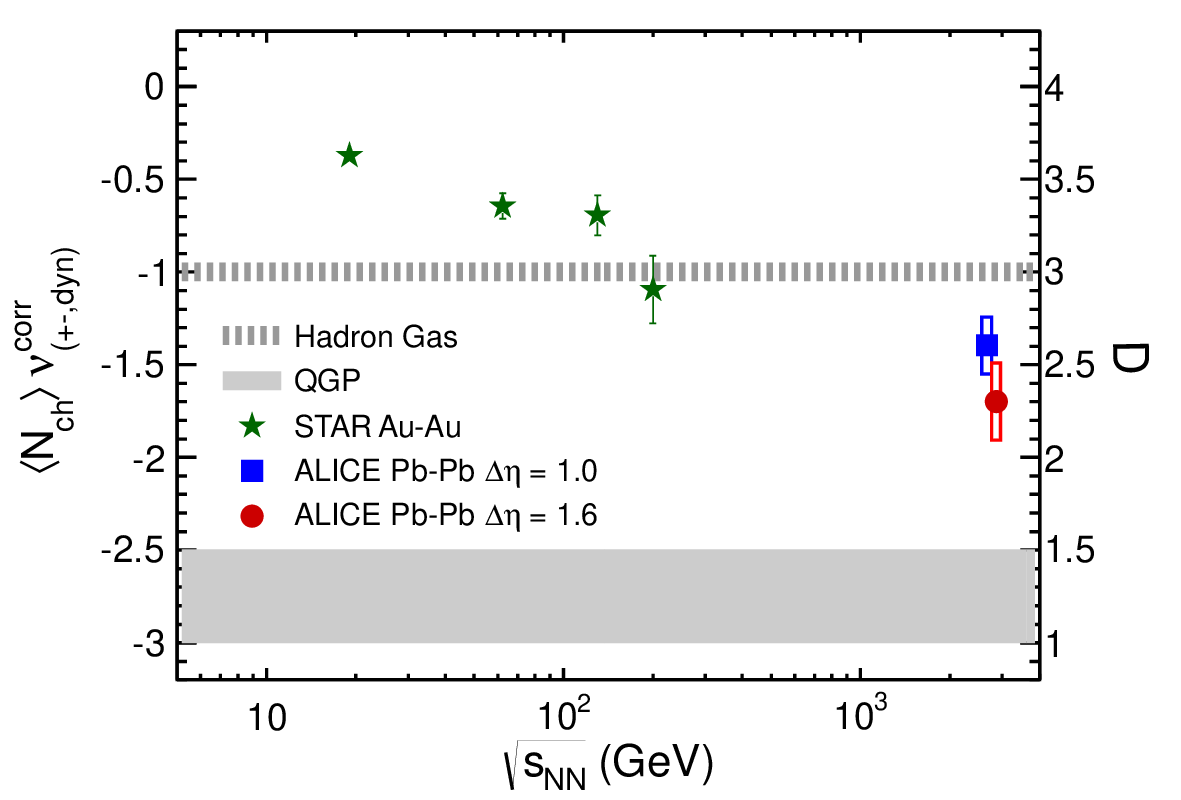}
\caption{  
Energy dependence of the net--charge fluctuations, 
measured in terms of \nchnudyn~(left axis) and $D$ (right axis), for the top central collisions.  
The results from the STAR~\cite{Abelev09} and ALICE experiments are presented for $\Delta\eta=1$ 
after the correction for charge conservation. 
The ALICE result for $\Delta\eta=1.6$ is also shown.
Both statistical (error bar) and systematic (box) errors are plotted.
}
\label{fig_Edepend}
\end{center}
\end{figure}

The evolution of the net--charge fluctuations with beam energy can be
studied by combining the ALICE data with those of the STAR experiment~\cite{Abelev09} at RHIC. 
In Figure~\ref{fig_Edepend}, we present the values of
\nchnudyn~(left axis) and $D$ (right axis) for top
central collisions from ALICE at $\sqrt{s_{\rm NN}}=2.76$~TeV
and for STAR, \mbox{Au--Au} collisions at four different energies.
The ALICE data points correspond to $\Delta \eta = 1$ and 1.6,
whereas for STAR, the values for $\Delta \eta = 1$ are shown.
For the STAR data, $(dN_{\rm ch}/d\eta)\nu^{\rm corr}_{(+-,{\rm dyn})}$ are plotted instead
of \nchnudyn, as $dN_{\rm ch}/d\eta$ values are approximately 
equal to $\langle N_{\rm ch} \rangle$ for $\Delta \eta = 1$ at central rapidity. 
The theoretical predictions for a HG and a QGP are indicated in the figure.
In the absence of any dynamic model, these predictions do not
    have dependence on the beam energy. 

Figure~\ref{fig_Edepend} shows a monotonic decrease in the
magnitude of the net--charge fluctuations with increasing beam energy. 
For the top RHIC energy of $\sqrt{s_{\rm NN}}=200$~GeV, the measured
value of fluctuation is observed to be close to the HG prediction,
whereas at lower energy, the results are above the HG value.
At $\sqrt{s_{\rm NN}}=2.76~TeV$, 
we observe significantly lower fluctuations compared to
those of lower energies. 

In summary, we have presented the first measurements of dynamic net--charge 
fluctuations at the LHC in \mbox{Pb--Pb} collisions at $\sqrt{s_{\rm NN}}=2.76$~TeV 
in terms of \nudyn, and their corrected values, \nudyncorr~(corrected 
for charge conservation and
finite acceptance effect). The results for \mbox{pp} collisions at the
same center--of--mass energy are found to be in agreement with hadron
gas prediction.
The values of \nudyn~and \nudyncorr are seen to be negative in all
cases, indicating the
dominance of the correlation of positive and negative charges. 
A decrease in fluctuations is observed while going from peripheral to
central collisions.
The $D$--measure, which gives the charge fluctuations per entropy, is
calculated from \nudyncorr~and from the measured average charged particle multiplicity.
A decreasing trend of $D$ is observed in going from peripheral to central
collisions. Model studies indicate that the presence of radial flow may not be the
cause of this decrease. The dissipation of signal during the
evolution of the fireball from the hadronization to freeze-out
has been estimated by fitting  $D$ as a function of the \deltaeta~window.
The extrapolation of the fit function yields the asymptotic value of
$D$, which is not very different
from the measurement at $\Delta \eta = 1.6$. 
The beam energy dependence of charge fluctuations has been studied by
comparing the ALICE data with those from the STAR experiment at RHIC for
\mbox{Au--Au} collisions at four energies.
A monotonic decrease in the value
of $D$, measured at $\Delta \eta = 1$, has been observed. The STAR data
points at RHIC top energy are close to the prediction for a hadron gas.
This may be due to the fact that the fluctuation may be not strong enough to be measured or 
because of the dilution of fluctuation during the evolution process.
The fluctuations at $\sqrt{s_{\rm NN}}=2.76~TeV$ for $\Delta \eta =
1.0$ are below the measurements at RHIC. 
These data points show an
additional decrease at $\Delta \eta = 1.6$, where
$D$ turns out to be
$2.3 \pm 0.02 {\rm (stat.)} \pm 0.21 {\rm  (sys.)}$ 
for top central collisions. The fluctuations at the LHC energy might
also have been diluted because of various effects as discussed earlier.
As these fluctuations are smaller than the
theoretical expectations for a HG, and show for the first time at 
the LHC energy a clear tendency towards expectations from a QGP, we may 
infer that they have their origin in the QGP phase.
Dynamical model calculations are needed to better understand the results.

\vspace{1 cm}
The ALICE collaboration would like to thank all its engineers and technicians for their invaluable contributions to the construction of the experiment and the CERN accelerator teams for the outstanding performance of the LHC complex.
\\
The ALICE collaboration acknowledges the following funding agencies for their support in building and
running the ALICE detector:
 \\
State Committee of Science, Calouste Gulbenkian Foundation from
Lisbon and Swiss Fonds Kidagan, Armenia;
 \\
Conselho Nacional de Desenvolvimento Cient\'{\i}fico e Tecnol\'{o}gico (CNPq), Financiadora de Estudos e Projetos (FINEP),
Funda\c{c}\~{a}o de Amparo \`{a} Pesquisa do Estado de S\~{a}o Paulo (FAPESP);
 \\
National Natural Science Foundation of China (NSFC), the Chinese Ministry of Education (CMOE)
and the Ministry of Science and Technology of China (MSTC);
 \\
Ministry of Education and Youth of the Czech Republic;
 \\
Danish Natural Science Research Council, the Carlsberg Foundation and the Danish National Research Foundation;
 \\
The European Research Council under the European Community's Seventh Framework Programme;
 \\
Helsinki Institute of Physics and the Academy of Finland;
 \\
French CNRS-IN2P3, the `Region Pays de Loire', `Region Alsace', `Region Auvergne' and CEA, France;
 \\
German BMBF and the Helmholtz Association;
\\
General Secretariat for Research and Technology, Ministry of
Development, Greece;
\\
Hungarian OTKA and National Office for Research and Technology (NKTH);
 \\
Department of Atomic Energy and Department of Science and Technology of the Government of India;
 \\
Istituto Nazionale di Fisica Nucleare (INFN) and Centro Fermi -
Museo Storico della Fisica e Centro Studi e Ricerche "Enrico
Fermi", Italy;
 \\
MEXT Grant-in-Aid for Specially Promoted Research, Ja\-pan;
 \\
Joint Institute for Nuclear Research, Dubna;
 \\
National Research Foundation of Korea (NRF);
 \\
CONACYT, DGAPA, M\'{e}xico, ALFA-EC and the HELEN Program (High-Energy physics Latin-American--European Network);
 \\
Stichting voor Fundamenteel Onderzoek der Materie (FOM) and the Nederlandse Organisatie voor Wetenschappelijk Onderzoek (NWO), Netherlands;
 \\
Research Council of Norway (NFR);
 \\
Polish Ministry of Science and Higher Education;
 \\
National Authority for Scientific Research - NASR (Autoritatea Na\c{t}ional\u{a} pentru Cercetare \c{S}tiin\c{t}ific\u{a} - ANCS);
 \\
Ministry of Education and Science of Russian Federation,
International Science and Technology Center, Russian Academy of
Sciences, Russian Federal Agency of Atomic Energy, Russian Federal
Agency for Science and Innovations and CERN-INTAS;
 \\
Ministry of Education of Slovakia;
 \\
Department of Science and Technology, South Africa;
 \\
CIEMAT, EELA, Ministerio de Educaci\'{o}n y Ciencia of Spain, Xunta de Galicia (Conseller\'{\i}a de Educaci\'{o}n),
CEA\-DEN, Cubaenerg\'{\i}a, Cuba, and IAEA (International Atomic Energy Agency);
 \\
Swedish Research Council (VR) and Knut $\&$ Alice Wallenberg
Foundation (KAW);
 \\
Ukraine Ministry of Education and Science;
 \\
United Kingdom Science and Technology Facilities Council (STFC);
 \\
The United States Department of Energy, the United States National
Science Foundation, the State of Texas, and the State of Ohio.



\begin{thebibliography}{99}
\bibitem{Aamodt08}     K.~Aamodt {\em et al.} [ALICE Collaboration],
  JINST {\bf 3}, S08002 (2008).
\bibitem{JeonKoch00}   S.~Jeon, V.~Koch, Phys. Rev. Lett. {\bf 85},  2076 (2000).
\bibitem{asakawa} Masayuki Asakawa, Ulrich Heinz, and Berndt Muller, Phys. Rev. Lett. {\bf 85},  2072 (2000).
\bibitem{dtilde} M.~Bleicher, S.~Jeon, and V.~Koch Phys. Rev. C {\bf
      62}, 061902 (2002).
\bibitem{JeonKochReview} Sangyong Jeon and Volker Koch, In
 Quark--Gluon--Plasma 3, Ed. R.C. Hwa and X.N. Wang, 430 (2004); arXiv:hep-ph/0304012v1.
\bibitem{JeonKoch99}   S.~Jeon, V.~Koch, Phys. Rev. Lett. {\bf 83},
5435 (1999).
\bibitem{Asakawa00}    M.~Asakawa, U.~Heinz, B.~Mueller,
Phys. Rev. Lett. {\bf 85},  2072 (2000).


\bibitem{Shuryak01}    E.~V.~Shuryak, M.~A.~Stephanov, Phys. Rev. C
  {\bf 63}, 064903 (2001).
\bibitem{AbdelAziz05}  M.~A.~Aziz, S.~Gavin, Phys. Rev. C {\bf 70}, 034905 (2004).
\bibitem{Voloshin} S. Voloshin, Phys. Lett. {\bf B632}, 490 (2006).
\bibitem{zaranek} J.~Zaranek, Phys. Rev. {\bf C66} 024905 (2002).
\bibitem{Abelev09}     B.~I.~Abelev {\em et al.} [STAR Collaboration],
  Phys. Rev. C {\bf 79}, 024906 (2009).
\bibitem{Adams03c}     J.~Adams {\em et al.} [STAR Collaboration],
  Phys. Rev. C {\bf 68},  044905 (2003).
\bibitem{CERES}        H.~Sako {\em et al.} [CERES/NA45
  Collaboration], Jour. Phys. G {\bf 30},  S1371 (2004).
\bibitem{NA49}           C.~Alt {\em et al.} [NA49 Collaboration], Phys. Rev. C
{\bf 70}, 064903 (2004).
\bibitem{AdcoxPRL89}   K.~Adcox {\em et al.} [PHENIX Collaboration],
  Phys. Rev. Lett. {\bf 89},  082301 (2002).
\bibitem{Pruneau02}    C.~Pruneau, S.~Gavin, S.~Voloshin, Phys. Rev. C
  {\bf 66}, 044904 (2002).
\bibitem{christiansen} P. Christiansen, E. Haslum, E. Stenlund,
  Phys. Rev. C {\bf 80}, 034903-1 (2009).
\bibitem{Nystrand}  J.~Nystrand, E.~Stenlund, H.~Tydesjo,
  Phys. Rev. \textbf{C68}, 034902 (2003).
\bibitem{TPC}  J.~Alme {\em et al.} [ALICE Collaboration],
  Nucl. Instr. Meth. {\bf A622}, 316 (2010).
\bibitem{ALICEcharged} K.~Aamodt {\em et. al.} [ALICE Collaboration],
 Phys. Rev. Lett. {\bf 106}, 032301 (2011).
\bibitem{Glauber} B.~Alver, M.~Baker, C.~Loizides, P.~Steinberg,
    arXiv:0805.4411 [nucl-ex] (2008).
\bibitem{miller}        M.L. Miller, K. Reygers, S.J. Sanders,
  P. Steinberg, Annu. Rev. Nucl. Part. Sci. {\bf 57}, 205 (2007).
\bibitem{toia}   A.~Toia {\em et al.} [ALICE Collaboration],
  Jour. Phys. {\bf G38}, 124007 (2011).
\bibitem{ALICE_ntotal}  
``Centrality dependence of the pseudorapidity density distribution for charged particles in Pb--Pb
   collisions at $\sqrt{s_{\rm NN}}=2.76$ TeV'',
ALICE Collaboration, to be published.
\bibitem{Ref:Pythia} T. Sjostrand and P. Skands, Eur. Phys. J. {\bf
    C39} 129 (2005).
\bibitem{Ref:Hijing} M.~Gyulassy and X.~N.~Wang,
   Comput. Phys. Commun. {\bf 83}, 307 (1994); 
W.-T. Deng, X.-N. Wang, and R. Xu, (2010),
arXiv:1008.1841 [hep-ph].
\bibitem{afterburner} E Cuautle and G Paic, Jour. of Phys. {\bf G35},
  075103 (2008).

\end{thebibliography}
\end{document}